# Can we predict interface dipoles based on molecular properties?

*Johannes J. Cartus, Andreas Jeindl, Oliver T. Hofmann\**

Institute of Solid State Physics, Graz University of Technology, Petersgasse 16/II, 8010 Graz, Austria

We apply high-throughput DFT calculations and symbolic regression to hybrid inorganic/organic interfaces with the intent to extract physically meaningful correlations between the adsorption-induced work function modifications and the properties of the constituents. We separately investigate two cases: Hypothetical, free standing self-assembled monolayers with a large intrinsic dipole moment, and metal-organic interfaces with a large charge-transfer induced dipole. For the former we find – without notable prior assumptions – the Topping model, as expected from literature. For the latter, highly accurate correlations are found, which are, however, clearly unphysical.

# Introduction

The level alignment of metal-organic interfaces has been subject to much attention from both fundamental[1–4] and engineering research, especially in the context of organic electronics.[5–8] Suboptimal choices in the design of the interface materials can lead to poor device performance, e.g. due to electrical resistances caused by large charge injection barriers.[5] However, these injection barriers, which depend on the offset between the metal's Fermi energy and the molecular levels,[2] can be optimized by adsorbing a so-called charge injection layer onto the metal. These layers change the level alignment due to the emergence of an adsorption-induced potential jump[1–4] (often termed "interface dipole"), $\Delta\Phi$.

Presently, $\Delta\Phi$ must be determined separately for every substrate/adsorbate combination, either experimentally or via first-principles calculations, but both options are expensive and laborious.[9] A prediction – or at least a solid estimate – of $\Delta\Phi$ based solely on properties of the isolated adsorbate and substrate would significantly speed up the optimization process for inorganic/organic interfaces. However, although there are several (often conflicting) models that relate molecular properties to $\Delta\Phi$ (such as the induced density of interface states model,[10–13] the integer charge transfer model,[14–16] or pinning on the molecular LUMO[17–21]), an explicit expression describing the interface-dipole via the properties of the constituents has yet to be put forth. In fact, it is not yet clear whether such an expression can be formulated based solely on properties of the interface constituents at all.

In this work, we attempt to extract an analytic expression by a combination of high-throughput first-principles calculations and symbolic regression. In its most simple formulation, symbolic regression takes a number of input properties (e.g., the molecular dipole moment, the ionization energy, see below) and combines them via mathematical operators (e.g. multiplication, exponentiation, etc.) into more complex equations (i.e., analytic models). These expressions are then fitted against a target quantity (e.g., ΔΦ). Ideally, the best-fitting models correspond to the "natural laws", that govern the physics underlying the data.[22,23] This approach can be seen as a one-dimensional variant of the SISSO method[24] (see Supporting Information). However, this additional complexity is not necessarily required – or helpful - for detecting physical relationship (see Supporting Information).

When studying the interface-dipole-induced work function change, it has become customary to dissect it into two components:[18,25] the contribution that arises from the bonding to the substrate, $\Delta\Phi^{Bond}$, and the jump of the electrostatic potential that would be induced by the adsorbate alone, $\Delta\Phi^{Mol}$:

$$\Delta\Phi = \Delta\Phi^{Mol} + \Delta\Phi^{Bond} \tag{1}$$

Since it is difficult, if not impossible, for symbolic regression to identify two different effects of similar magnitude (explanation see Supporting Information), here we aim to obtain analytical models for $\Delta\Phi^{Mol}$ and $\Delta\Phi^{Bond}$ separately.

## Results and Discussion

**Adsorbate dipole.** Starting with the adsorbate-dipole-induced potential jump $\Delta\Phi^{Mol}$, we select 6 different heteroaromatic molecules (shown in Figure 1). We designed several planar hetero aromatics with substantial in-plane dipole moments through specific substitution of halogens and nitrogen. From these we selected the 6 molecules with the largest electron affinity, which will be useful for the investigation of $\Delta\Phi^{Bond}$ later in this work. We then created hypothetical, free-standing self-assembled monolayers by placing the molecules with their intrinsic dipole moment aligned in z-direction in various unit cells with various side lengths (12.5-30 Å) and angles (45, 60, 75, 90°).

The combination of molecules and unit cells yields 360 different systems. For these systems, we obtain $\Delta\Phi^{Mol}$ by performing dispersion-corrected DFT calculations with FHI-aims[26] using the PBE exchange-correction functional[27] and a dipole correction[28] (further details are given in the Methods section).

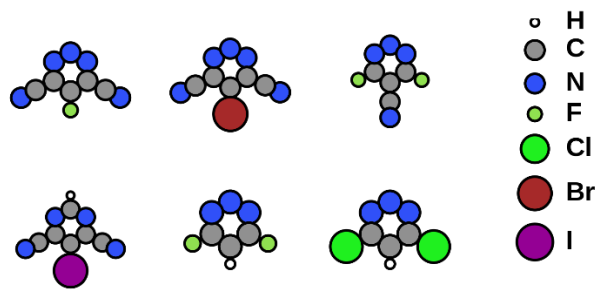

**Figure 1:** The 6 heteroaromatic molecules we used to build free-standing self-assembled monolayers.

To extract analytic models for $\Delta\Phi^{Mol}$, our symbolic regression algorithm takes various properties of the interface or its constituents as input. We then combine these input properties via mathematical operations to build analytical expressions. To keep the number of possible expressions tractable, we adhere to the following protocol: In a first step we exhaustively create products of up to three input parameters and their reciprocals, i.e. we create expressions of the form $F(x,y,z) = x^a y^b z^c$ with three different input parameters $x, y, z$ and exponents $a, b, c \in \{-1, 0, 1\}$. In the second step, we create additional expressions by applying the nonlinear mapping $F'(F_i, F_j) = F_i/(F_j + 1)$ to all possible pairs of expressions, with the restriction that the factors x/y/z in $F_i$ and $F_j$ can only differ in a single input parameter. Finally, all created expressions are evaluated using the input parameter values from the systems in the dataset. For each analytical expression, a linear fit against $\Delta\Phi^{Mol}$ is performed, and the best performing fit (in terms of its root-mean-square-error, RMSE) is reported.

Since the resulting set of analytical expressions grows very fast with the number of input parameters, a thoughtful selection is required. Here, we use the following properties for the isolated molecules in gas phase as input parameters: the orbital energy of the highest occupied molecular orbital ($\varepsilon_{HOMO}$) and of the lowest unoccupied molecular orbital ($\varepsilon_{LUMO}$), the ionization potential (IP) and the electron affinity (EA) via the $\Delta$SCF-approach (see Methods section), the molecular dipole moment µ and the molecular polarizability α along the direction of the dipole moment. In addition, we provide the lengths of the unit cell vectors (a,b), minimum distance between two atoms ($d_{min}$), and $C_\Sigma$, the infinite sum of cubed reciprocal distances from a dipole to all its neighbors as geometry-dependent input parameters. The latter often appears in the electrostatic description of collective electric fields of dipoles.[29] A compilation of all used parameters is provided in Table 1.

**Table 1:** Compilation of input parameters used to construct the candidate analytical expressions for $\Delta\Phi^{Mol}$.

| Name | Description |
| --- | --- |
| $a, b$ | Unit cell side lengths |
| $d_{min}$ | Minimum distance between atoms of periodic replicas of adsorbate |
| $\rho$ | Dipole density (number of molecules per area) |
| $C_\Sigma$ | Infinite sum of cubed reciprocal distances $r_i$ from a molecule to all its neighbors: $C_\Sigma = \sum_i r_i^{-3}$. |
| $\varepsilon_{HOMO}, \varepsilon_{LUMO}$ | Orbital energies of adsorbate molecule |

| IP, EA | Vertical ionization potential (IP) and electron affinity (EA) |
| --- | --- |
| $\mu_z$ | z-component of the single adsorbate dipole moment in vacuum |
| $\alpha_{zz}$ | Molecular polarizability along the direction of the dipole moment. |

For our systems, the expression with the lowest RMSE (3.3 meV) is found to be

$$\Delta\Phi^{Mol} \propto \frac{\mu_z \rho}{(\alpha_{zz} C_\Sigma + 1)} \quad (2)$$

The excellent agreement between the prediction of $\Delta\Phi^{Mol}$ via equation 2 and the "true" values originally obtained by DFT is displayed in Figure 2a. Equation 2 is exactly the Topping-equation, which is expected from classical electrostatics[29–31] and was previously suggested by numerous other theoretical[32–34] and experimental[31,35] works.

While finding the Topping model from our data shows the validity of our approach, it is important to emphasize that this success is by no means guaranteed. Obviously, we could only find equation 2 because we allowed for the non-linear mathematical operation and because we provided $C_\Sigma$ as input parameter. Neither of these would necessarily be intuitive, but without either of these, we would only obtain physically incorrect solutions. Interestingly, when including additional systems that are too densely packed (i.e., when the point-dipole approximation underlying the Topping model[29] starts to break down), some of these unphysical models exhibit even lower RMSE values (i.e., perform even better) than the physically correct expression. Nevertheless, under the correct boundary conditions, the "correct" physical picture can be accurately obtained from our data.

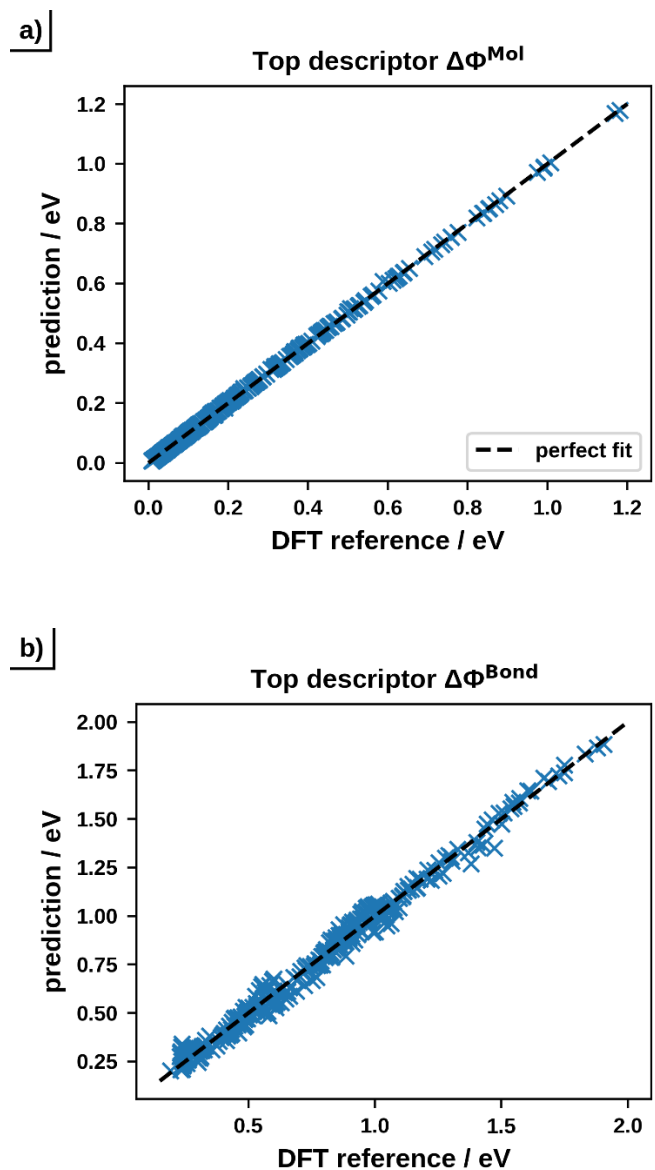

**Figure 2:** Prediction of the a) $\Delta\Phi^{Mol}$ and b) $\Delta\Phi^{Bond}$ versus the DFT-calculated reference data. The dashed line marks a hypothetical perfect fit.

**Bond dipole.** As second step, we turn to $\Delta\Phi^{Bond}$. This term contains all effects of the interface dipole arising from the interaction of the adsorbate with the substrate, such as charge transfer, Pauli pushback, formation of covalent bonds, etc.[1–4] In this work, we focus on charge transfer only, because it is a) relatively straightforward to separate from the other effects and because b) one would expect that the molecular properties that govern it can be easily tuned via chemical modifications of the adsorbate.

To increase the diversity in our dataset (compared to the previous section), here we use a total of 28 different heteroaromatic molecules, consisting of the 6 molecules used before plus 22 more that are based on naphthalene as a backbone to allow for more varied molecular properties (details see Supporting Information). A challenge when considering $\Delta\Phi^{Bond}$ is that it is known to depend strongly on the geometry the adsorbate assumes. For instance, whether a molecule adsorbs flat-lying or upright standing can change $\Delta\Phi^{Bond}$ by more than 1eV,[36–38] due to the associated change in the molecule's ionization energies.[39] Similarly, $\Delta\Phi^{Bond}$ is strongly affected by the molecular coverage.[33,40,41] However, on metal substrates most molecules adsorb in a flat-lying geometry.[42] We assume that this is also the case for all our molecules. This has the further advantage that their molecular dipole moment is parallel to the xy-plane, such that it does not contribute to $\Delta\Phi$. To remove any coverage-dependent effects, we use the same supercell, i.e. adsorbate density, throughout. This geometry is an Ag(111) surface slab with 5 layers and a surface area of 5x5 Ag atoms. The large supercell ensures that there is only very little interaction between adjacent molecules. To focus on charge transfer only, we use hypothetical interfaces where the adsorption height is sufficiently large to inhibit wave-function overlap between the substrate and the adsorbate, thereby switching off any contributions from Pauli Pushback or covalent bonds. In practice, we adsorb molecules at distances between 7 and 100 Å above metal slabs made of Ag, Al, In, Mg, and Na. In total, this results in 323 different interface systems to evaluate our expressions on.

Also here, we must select suitable input parameters for the symbolic regression approach. We start by taking the ones suggested by the various models in literature:

- The induced density of interface states implies a relation to the density of states at the Fermi level (a substrate property),[10,11] which we also include here.
- The so-called integer charge transfer-model postulates that the charge-transfer occurs via polaronic levels.[14–16] While polarons are a crystal quantity and their relation to purely molecular properties is not perfectly clear, here we incorporate them approximately as the molecule's vertical electron affinity EA and the relaxation energy $E_{relax}$. The latter refers to the difference between the energy of the singly charged molecule in the geometry of the neutral molecule and when it is fully optimized, i.e. its internal reorganization energy. Since we are interested in work function changes, we use EA relative to the substrates' work function.
- Many previous theoretical calculations imply that $\Delta\Phi^{Bond}$ is determined by the difference between the frontier orbital energies and the Fermi energy.[18,19]

We also add the polarizability of the molecule perpendicular to the aromatic plane, $\alpha_{zz}$, and the HOMO-LUMO gap, since it is a common measure for the reactivity of a molecule.[43] . Furthermore, it is conceivable that image-charge effects play a role. As geometric properties, we therefore include the height of the molecule above the substrate's image plane position (details see Supporting Information) and above the topmost layer as input parameters. A comprehensive compilation of input parameters is given in Table 2. Otherwise, we construct expressions as we did for $\Delta\Phi^{Mol}$, i.e. we build products F of up to 3 input parameters (or their reciprocals), and create additional features via the non-linear mapping F' = $F_i/(F_j +1)$ using the same conditions as above.

**Table 2:** Compilation of input parameters used to find the expression for $\Delta\Phi^{Bond}$.

| Name | Description |
|---|---|
| $\varepsilon_{LUMO} - E_F$ | Difference of the LUMO of the adsorbate and the Fermi energy of the substrate. |
| $\varepsilon_{HOMO} - E_F$ | Difference of the HOMO of the adsorbate and the Fermi energy of the substrate. |
| $EA - \Phi$ | Difference of the EA of the adsorbate and the work function of the substrate. |
| $IP - \Phi$ | Difference of the IP of the adsorbate and the work function of the substrate. |
| $E_{relax}$ | Relaxation energy (see main text). |
| $DOS(E_F)$ | Density of states of the pristine substrate at the Fermi energy. |
| $h - z_{im}$ | Adsorption height of the molecule w.r.t the image plane position. |
| $h$ | Adsorption height of the molecule w.r.t the uppermost substrate layer. |
| $\varepsilon_{LUMO} - \varepsilon_{HOMO}$ | HOMO-LUMO gap. |
| $\alpha_{zz}$ | zz-component of the polarizability tensor of the adsorbate. |

For the bond dipole, we find the best performing expression to be:

$$\Delta\Phi^{Bond} \propto (\varepsilon_{LUMO} - E_F) \cdot \frac{h}{h - z_{im}} \cdot \frac{1}{1 + \left(\frac{IP - \Phi}{h}\right)} \quad (3)$$

With an RMSE of 38 meV, it performs reasonably well (see Figure 2b), even if the RMSE is one order of magnitude worse than the Topping model for the $\Delta\Phi^{Mol}$.

Inspection of equation 3 reveals that it is almost exclusively dominated by the term $(\varepsilon_{LUMO} - E_F)$. The second term, $h/(h-z_{im})$, is always larger than but close to 1 (between 1.0 and 1.5), because $h \gg z_{im}$ for the majority of systems in our dataset. Conversely, the third term is also always close to one, but smaller (between 0.5 and 1.0), because $IP-\Phi \ll h$ for all systems. Since these two terms also counteract each other (both get closer to 1 as h increases), the $\Delta\Phi^{Bond}$ values predicted by eq.

3 scatter only very little around $\varepsilon_{LUMO} - E_F$. This may, in principle, indicate that our computations yield noisy results for $\Delta\Phi^{Bond}$. However, our SCF procedure converged $\Delta\Phi$ to $10^{-4}$ eV (see Methods section), i.e., much too tight to allow for noise of this magnitude. Furthermore, we note that $\varepsilon_{LUMO}$ is an auxiliary quantity from DFT. There is some debate about when orbital energies correspond to observables (e.g., photoemission resonances).[44] However, they always do depend on the chosen exchange-correlation functional, i.e. the chosen theoretical method. In an earlier work, we have shown that there is no direct proportionality between ($\varepsilon_{LUMO} - E_F$) and $\Delta\Phi^{Bond}$ when going, e.g., from semi-local to hybrid functionals.[19] In other words, in salient contrast to the Topping model found in equation 2, equation 3 fails to extract the physics that governs the charge transfer at the interface. (Rather, it merely shows an excellent correlation.)

There are multiple possible reasons for this. In principle, it would be conceivable that some of our data is faulty. However, we can readily extract physical relationships for other physical properties, such as the adsorption energy (see Supporting Information), which attests to the fidelity of our results. Another possible explanation would be that we do not include the correct input parameters and mathematical operations, or that we do not allow for sufficiently complex expressions. However, also various other, additional input parameters and more varied exponents and functions fail to yield physically meaningful results (see Supporting Information). While this is no proof that we just did not include the "right" ingredients (such a proof is fundamentally impossible), it seems unlikely that the correct relation is an expression that is even more complex than what we already found for equation 3. Finally, we must face the hypothesis that our data – being synthetic, computed data with an approximate theory – just does not reproduces the underlying physics with sufficient accuracy.

Indeed, the PBE functional is known to have certain issues when describing charge-transfer systems. For example, when dissociating $H_2^+$ (i.e., placing the two H-cores far away from each other), it yields the unphysical solution of two protons with half an electron each, instead of a neutral H and a positively charged $H^+$ atom. Also here, we find that the molecules, even far above the surface (and thus completely unhybridized with it) are fractionally charged. In principle, a physically more correct solution could be a mixture of charged and uncharged molecules, i.e. integer charge transfer. This could only be obtained by employing large supercells in combination with specifically tuned hybrid functionals.[19] However, the optimal functional would have to be determined separately for each system,[45] which incurs computational costs that are presently intractable. At the same time, it is not clear whether this would even solve the issue: Since the interface dipole depends on the average amount of charge transferred per area, not its distribution, a computation with hybrid functionals may not yield more accurate values. A further, related problem is the self-interaction error of PBE. This effect not only causes the well-known underestimation of the band gap, but it also makes the energies of the orbitals (including the LUMO) dependent on their occupation. This shift of the orbital energy may be superimposed to the shift of the orbital energy induced by $\Delta\Phi^{Bond}$, making it impossible for symbolic regression to extract either effect.

## Conclusion

In summary, we attempted to extract the physics that govern the formation of interface dipoles at organic/inorganic interfaces. To that aim, we computed large datasets using semi-local density functional theory and applied symbolic regression to obtain functional relationships between properties of the molecules (and the substrate) and $\Delta\Phi$. The approach was successful for the contribution of the molecular dipole, yielding the well-known Topping-model. Conversely, for the charge-transfer contribution we obtained a clearly unphysical result that depends on a DFT-quantity rather than a molecular property. We tentatively attribute the failure to extract a clear physical relationship to shortcomings of the underlying method (PBE). Despite the generally outstanding performance of dispersion-corrected PBE calculations for interfaces,[46–49] this advises that caution should be taken when computing interfaces with a notable charge transfer character.

Moreover, the difficulty to extract the relevant physics for charge transfer even with a very large dataset and an extensive combination of molecular and substrate parameters shows that the design of inorganic-organic interfaces with a predefined level alignment is non-trivial and will continue to be so. Even when minimizing the impact of the adsorbate geometry (which is extremely difficult to predict in the first place), and when simplifying the problem by avoiding quantum mechanical interactions (such as Pauli Pushback and covalent bonds) as much as possible, we can only extract empirical correlations so far. For a comprehensive understanding and description of all effects at the interface, evidently much larger, more sophisticated data sets are still needed.

## Methods

All density functional theory (DFT) calculations mentioned in the paper were performed using FHI-aims.[26] This code allows to employ both open and periodic boundary conditions, i.e. individual molecules and interfaces can be treated on the same footing. For all systems, we used tight basis sets and numerical defaults as shipped with release 201103. The Perdew-Burke-Ernzerhof (PBE)[50] exchange-correlation functional was used together with the vdW-TS[51] dispersion correction.

To obtain the properties of the individual molecules, we performed calculations with open boundary conditions. The geometry of the (charge neutral) molecules was fully relaxed until the remaining forces on each atom fell below 0.01 eV/Å. From the optimized geometry, we extracted the orbital energies of the HOMO and LUMO, the molecular dipole moment and the polarizability (via density functional perturbation theory[52]). Furthermore, we calculated the first and the second vertical ionization energy (IE) and electron affinity (EA) using the so-called ΔSCF-approach.[53,54] There, these energies are given as the energy difference between the singly charged and the uncharged molecule while keeping the geometry of the neutral molecule. The singly charged molecules are calculated spin-polarized (which is not necessary for the neutral molecules). We employed a Gaussian occupation scheme with a broadening of 0.01 eV.

All other calculations (free-standing monolayer, metal-organic interfaces, bare metals) were performed with periodic boundary conditions. Calculations for the bare substrate as well as for metal-organic interfaces were done using 5 metal layers. We employed a repeated slab approach to emulate 2D periodicity. The unit cell heights were chosen so that the vacuum amounts to at least 50 Å. To electrostatically decouple the periodic replica in z-direction, we used a dipole correction.[28] The SCF algorithm was repeated until total energies in subsequent iterations differed by less than $10^{-5}$ eV and electron densities differed by less than $10^{-3}$ electrons. Furthermore, we ensured for all calculations that the change in work function is converged to better than $10^{-4}$ eV between subsequent SCF iterations, as suggested by best practices.[49] We employed a generalized Monkhorst-Pack k-point grid[55,56] that corresponds to 50x50x1 k-points for the primitive substrate cells. Furthermore, a Gaussian occupation scheme with a broadening of 0.1 eV was used.

# Supporting Information

## Our method vs SISSO

SISSO[1] works in an iterative, two-stage process to find the expression that best describes a target quantity. Expressions are generated and evaluated as they are in our approach; see main text. The first stage preselects expressions based on correlation with the target and the second stage builds linear models out of the preselected expressions. In the first iteration it finds the single best analytic expression to describe the target as a linear model. This is equivalent to our method. In the next iteration the residual of the previous linear model is calculated, and expressions are preselected based on correlation with the residual. Again, linear regression is used to produce a linear model describing the target (now with 2 terms). This is repeated until the final iteration is reached (the linear model now has as many terms as iterations were performed). The number of iterations/terms is a user specified parameter and called "dimension of the descriptor".

The preselection based on the residual of previous iteration is what makes the method impractical for our uses. Say the target physics we want to describe is given by the sum of two terms. Intuitively, it seems logical to aim for a two-dimensional descriptor. However, as neither of the individual terms correlates well with the target data, neither of them is preselected in the first stage of the SISSO iterations. This makes it impossible to find a two-dimensional descriptor describing the data. Conversely, if the sum of the two terms is considered as an expression directly, it will most probably be found because the expression will correlate well with the target. This corresponds to a one-dimensional descriptor for the SISSO method, thus showing that additional dimensions are not additionally useful.

## Finding concurrent effects at the same time

Our method relies heavily on of sensible expressions being better correlated with the target than other, unphysical expressions. The threshold for necessary correlation is very high, as is also demonstrated by the results in the main text. There, the presented top ranked expressions show Pearson correlation coefficients[2] of greater than 0.95; this includes also clearly unphysical expressions. Sensible expressions are thus most easily found if the corresponding physics is well represented in the data. However, when there are multiple, concurrent physical effects present in the data it obviously becomes harder to capture singular effects.

We demonstrate this with a 1-D example: assuming the physics of a problem at hand is given by a known 1-D function, one can see how well the known function correlates with the available data when we introduce a secondary effect as perturbation. Let the physics of interest be described by a sine function of an independent variable t. Let an additional, concurrent effect be described by a cosine. The available data shall then be given by

$$F_\alpha(t) = (1 - \alpha) \sin(t) + \alpha \cos(t). \tag{1}$$

$\alpha$ is the mixing parameter which determines how much of the concurrent effect is mixed into the target (i.e. how strong the perturbation is compared to the main effect). For $t \in [0, 10]$ we have plotted the functions below in Fig. 1.

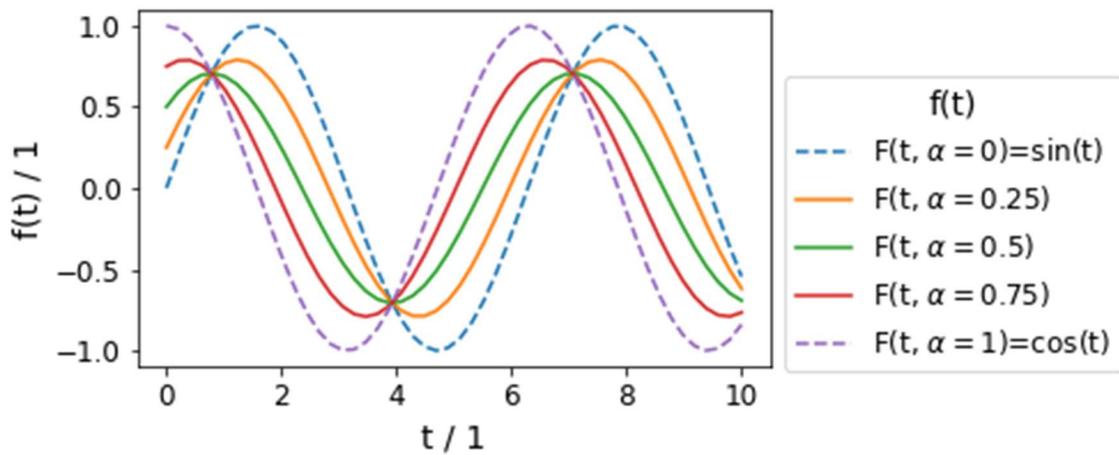

Figure 1: Various functions f(t) of the independent variable t. The values of $F_\alpha(t)$ represent available data used to calculate correlation coefficients.

Even though sine, cosine and $F_\alpha$ are very similar in their functional form the correlation sin(t)/$F_\alpha$(t), measured via Pearson correlation coefficient $\rho$, decreases with increasing $\alpha$, until it vanishes (Fig. 2).

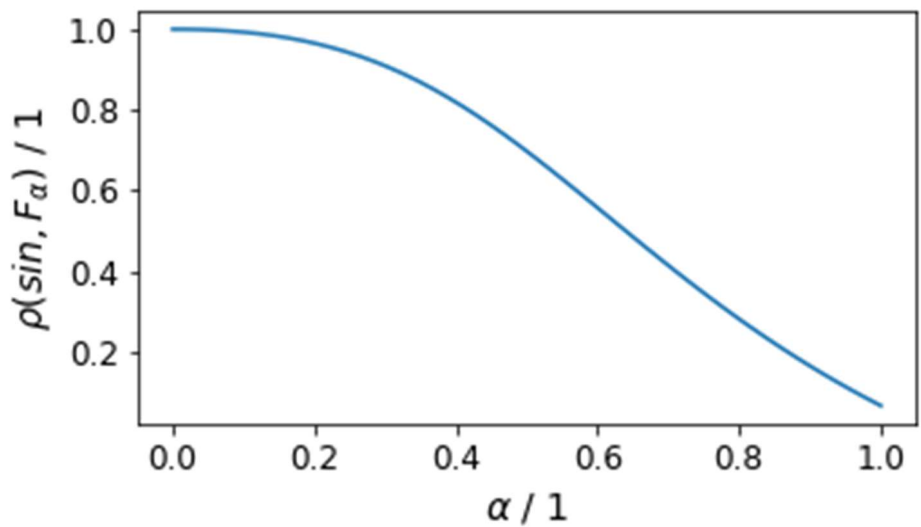

Figure 2: Correlation of sin(t) with $F_\alpha(t)$ for different values of $\alpha$, measured with Pearson correlation coefficient $\rho$ for t ∈ [0, 10].

This shows very nicely how a concurrent effect can interfere with detecting physics even in this very simplified example. In reality, one faces multiple concurrent effects simultaneously, which depend on multiple independent variables in addition to noise and systematic errors from the used method of measurement/computation.

## Naphthalene-derived heteroaromatic molecules

In addition to the 6 benzene-derived heteroaromatic molecules (Fig. 1, main text) we used the following 22 naphthalene base molecules as adsorbates in the metal-organic interfaces we study.

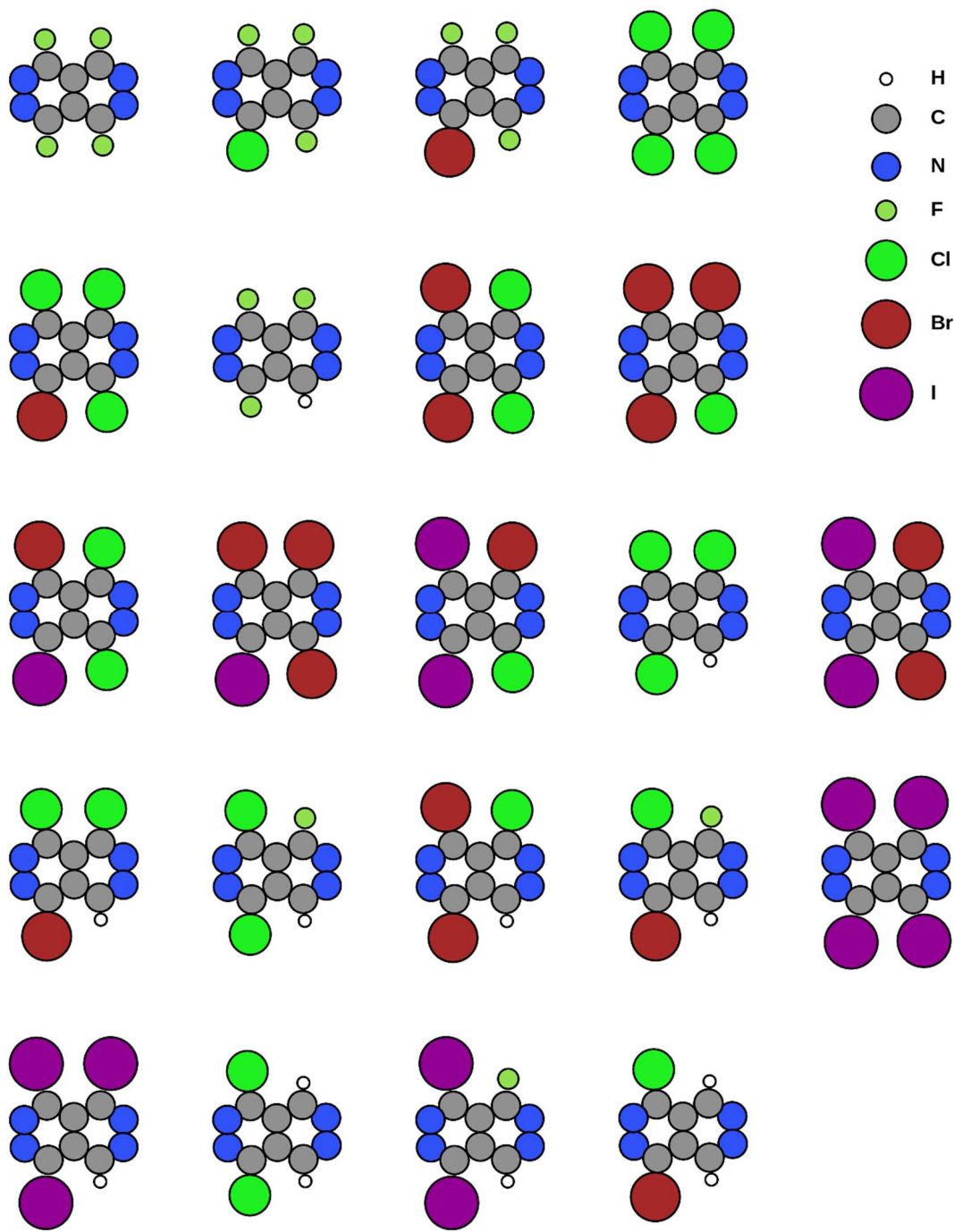

*Figure 3: Naphthalene derived heteroaromatic molecules used to build metal-organic interface systems.*

# Image plane positions for substrates

As explained in the main text, substrate polarization effects can be modelled with a classical image potential $V_{\text{im}(z)} = -\frac{1}{4(z-z_{\text{im}})}$. The introduction of a test charge close to the surface must yield a response of the electronic density (due to screening of the test charge). The center of mass of this change must correspond to $z_{\text{im}}$.[3] The calculated the electronic density for the pristine surface and the surface with a perturbation of positive charge of +0.01 electrons. In detail, we use a single atom with no basis functions and a modified core charge. This atom donates its "electron" (0.01 e[-]) into the substrate yielding a positive charge. The test charge is put 7 Å and 10 Å above the surface. For both heights, the center of mass of the disturbance is calculated. Finally, the average is used.

We applied this method to all substrates used in this work. The resulting image plane positions are listed in Table 1.

*Table 1: Image plane position for the used substrates.*

| Substrate slab, (111) surface | Image plane position / Å |
| --- | --- |
| Ag | 2.278 |
| Al | 2.885 |
| In | 2.863 |
| Mg | 2.278 |
| Na | 4.241 |

## Dataset sanity & alternative approach

In order to check the fidelity of our data we attempted to find an expression for the work function change with an alternative set of parameters. The input parameters we used are listed in Table 2. We generated expressions by multiplying up to 5 input parameters (raised to the powers {-1, 1, 2}) with each other. Here, we also include parameters like the charge Q of the adsorbate. We note, however, that such a parameter is per se unphysical, because it is not an observable itself.

*Table 2: Input parameters used to generate expressions for the adsorption energy*

| Name | Description |
| --- | --- |
| h - $z_{im}$ | Adsorption height of the molecule w.r.t the image plane position. |
| h | Adsorption height of the molecule w.r.t the uppermost substrate layer. |
| $\varepsilon_{HOMO}$, $\varepsilon_{LUMO}$ | HOMO and LUMO orbital energy of adsorbate. |
| $\partial \varepsilon_{LUMO} / \partial n$ | Change of LUMO orbital energy w.r.t occupation. |
| IP, EA | Ionization potential and electron affinity of adsorbate. |
| $EA^{2nd}$ | Second electron affinity. |
| $DOS(E_F)$ | Density of states of the pristine substrate at the Fermi energy. |
| $\Phi_0$ | Work function of the pristine substrate. |
| Q | Charge transferred from substrate to adsorbate (from Mulliken analysis). |

We find

$$\Delta \Phi \propto Q(z - z_{im}) \qquad (2)$$

as best performing expression with an RMSE of 23 meV. As this corresponds to the potential difference between the plates of plate capacitors with constant area (distance of the plates: $z$-$z_{im}$). Given the small RMSE and the very plausible expression, this result attests the sanity of our data.

Thus, the question arises if we could predict Q from input parameters of the interface constituents alone. Using the input quantities from Table 2 (except for Q itself) and allowing for products with up to 5 factors, all of which are powers of input parameters with the powers {-1, 1, 2}, we obtain

$$Q \propto \frac{EA \cdot IP^2}{h \cdot DOS(E_F) \cdot \Phi_0} \qquad (3)$$

as best expression. Unfortunately, the expression is clearly unphysical, making this approach unsuccessful as well.